\documentclass[draft]{agujournal2019}

\usepackage{url}
\usepackage{lineno}
\usepackage[inline]{trackchanges}
\usepackage{ragged2e}
\usepackage{microtype}
\usepackage{amsmath,amssymb}
\usepackage{graphicx}
\usepackage{booktabs}
\usepackage{siunitx}
\usepackage{bm}
\usepackage{apacite}
\usepackage{afterpage}

\graphicspath{{images/}}

\newcommand{\citep}[1]{\cite{#1}}

\newcommand{\citet}[1]{\citeA{#1}}

\newcommand{\vect}[1]{\bm{#1}}

\draftfalse
\journalname{JGR: Space Physics}

\begin{document}
\justifying
\setlength{\emergencystretch}{2em}

\title{Shrinking the Haystack: One-Class Machine Learning Detection of Magnetosheath Current Sheets in MMS Burst Data}

\authors{Deep Ghuge\affil{1,2}, Daniel J. Gershman\affil{2}, Vadim Uritsky\affil{1,2}, Julia E. Stawarz\affil{3}}
\affiliation{1}{Department of Physics, The Catholic University of America, Washington, DC, USA}
\affiliation{2}{NASA Goddard Space Flight Center, Greenbelt, Maryland, USA}
\affiliation{3}{School of Engineering, Physics, and Mathematics, Northumbria University, Newcastle upon Tyne, UK}
\correspondingauthor{Deep Ghuge}{ghuge@cua.edu}

\begin{keypoints}
\item We target the small, short current sheets that develop during magnetosheath turbulence in MMS burst-mode data
\item A minimum-variance gate and a one-class anomaly detector reduce \num{22775} burst windows to 270 candidates, a 98.8\% search-space compression
\item Manual inspection confirms 78\% of retained candidates are sheet-like or reconnection-like signatures across 15 magnetosheath intervals
\end{keypoints}

\begin{abstract}
We present a morphology-first framework for narrowing the search for magnetic-reconnection candidates in Magnetospheric Multiscale (MMS) burst-mode data. The target is the small, short, and frequently electron-only reconnecting current sheets that occur in turbulent magnetosheath plasma. The pipeline operates in two stages. A local frame-quality gate based on minimum-variance analysis first retains only windows whose current-sheet coordinates are well defined. A one-class Deep Support Vector Data Description neural network then scores those windows against a library of 3{,}000 physically calibrated synthetic current sheets generated by Monte Carlo from a single published reference event. Acceptance into the surrogate library is governed by the second-order structure function $S_{2}(\tau)$: a candidate is admitted only if its multi-scale fingerprint tracks that of the seed event inside a tolerance band, together with a small number of shape-based checks. This $S_{2}(\tau)$-anchored construction defines the in-class distribution directly from a well-understood reference event and sidesteps the absence of a curated negative class in turbulent magnetosheath data. Applied to 15 magnetosheath turbulence intervals from the literature (1.58~h of burst-mode coverage), the framework compresses \num{22775} sliding windows to 270 candidate detections (a 98.8\% reduction). Manual visual screening identifies 93 of these as candidate reconnection events and a further 118 as sheet-like, retaining 78\% of the queue for follow-up; the candidate-reconnection pool extends well beyond the 22 detections that overlap the published reconnection-event catalog used here as a sanity check. The framework is intended as the data-reduction stage of a broader reconnection-search workflow, offered here as an initial proof of concept before extending the one-class design to additional feature channels.
\end{abstract}

\section*{Plain Language Summary}
Magnetic reconnection releases stored magnetic energy, heats plasma, and accelerates particles. NASA's MMS spacecraft record short bursts of high-rate plasma data in which reconnection signatures may appear, but those events are brief and embedded in a noisy turbulent background, so finding them by hand is slow. We built a two-stage automated finder: a simple physics check first throws away windows of data whose local geometry is poorly defined, and a small machine-learning model then scores the remaining windows for current-sheet shape. The model is trained on 3{,}000 synthetic ``what-if'' versions of a single well-studied real event, generated so that they share the same statistical fingerprint as the original. Tested on 15 magnetosheath intervals from the literature, the system shrinks tens of thousands of candidate windows down to a few hundred, the majority of which pass a human reviewer as plausible current sheets, including some not previously catalogued. The tool is a candidate-finder, not a final classifier; deciding which candidates are reconnection still requires physical analysis.

\section{Introduction}
\label{sec:intro}

Magnetic reconnection converts magnetic energy into particle heating, bulk flows, and nonthermal acceleration, and it is the central transport mechanism in many space and astrophysical plasmas \citep{Yamada2010, Zweibel2009, Hesse2020}. In Earth's magnetosheath, reconnection occurs at thin current sheets that develop within turbulence, including the small electron-only mode reported by \citet{Phan2018}. These crossings are short, morphologically diverse, and frequently embedded in elevated background fluctuations, all of which complicate automated detection in MMS burst-mode data \citep{Burch2016, BurchScience2016, Torbert2018, Hubbert2022}.

A reasonable taxonomy of existing detection strategies has three branches. (1)~\textit{Threshold-based methods} flag candidates whenever a single-window scalar diagnostic exceeds a chosen value. A widely used example is the Partial Variance of Increments (PVI), the magnetic-field increment $|\Delta\vect{B}(t,\tau)|$ at lag $\tau$ normalised by the root-mean-square value of $|\Delta\vect{B}|$ over the surrounding interval, originally developed for solar-wind turbulence and also applied directly to MMS magnetosheath current sheets \citep{Greco2008, Servidio2011, Osman2014, Yordanova2020}. More elaborate variants combine several thresholds at once, typically requiring a $B_L$ reversal, a localised peak in $|\vect{J}|$ above a chosen value, a minimum magnetic-shear angle, and in some studies a multi-spacecraft timing-consistency check or a Wal\'en-style jet test \citep{Hou2021, Gingell2020}. Layering criteria reduces false positives at the cost of excluding genuine crossings that satisfy most but not all checks. These methods are interpretable and inexpensive, but the trigger depends on amplitude rather than shape, so a small but morphologically clean crossing on a quiet background can score below threshold while a large but morphologically diffuse fluctuation scores above it. (2)~\textit{Supervised machine-learning methods} train classifiers on labeled events, for example a multilayer perceptron trained on published electron-diffusion-region events \citep{Lenouvel2021} or a recurrent neural network trained on particle-in-cell reconnection simulations \citep{Waters2025}. A multilayer perceptron is a feed-forward neural network that maps a fixed set of input features to a class label through several fully connected layers; a recurrent neural network instead processes the waveform sequentially, carrying an internal state that lets it use temporal context along the crossing. Both inherit the labeling convention used to build the training set, and a curated, statistically representative negative class is hard to construct in turbulent magnetosheath data, where ``not-a-current-sheet'' is operationally vague. (3)~\textit{Unsupervised and anomaly-detection methods} avoid the labeling problem entirely \citep{Finley2024, Camporeale2019}, but most do not yield a calibrated, geometrically interpretable score by themselves.

We propose a fourth branch, distinct from all three above: the present method scores morphological similarity to a calibrated reference shape, rather than thresholding a single scalar diagnostic, training a classifier on labeled events, or flagging generic statistical outliers. A current-sheet crossing has a characteristic multi-scale signature in the second-order structure function $S_{2}(\tau)$ (formally defined in Equation~\ref{eq:s2def}), and the surrogate library is built so that the trained network is sensitive to that signature regardless of absolute amplitude. Two windows that look alike shape-wise to a human reviewer should score similarly even if their amplitudes differ by an order of magnitude, while two windows with similar peak $|dB/dt|$ but different shape should score differently; a pure-threshold method cannot make either distinction without retuning. The cost is computational, since the multi-scale fingerprint and the encoder evaluation require more work per window than a single threshold comparison.

A subtler problem cuts across all three branches: researchers do not fully agree on which morphologies count as ``reconnecting'' versus ``current sheet without reconnection,'' so boundary placement, multi-spacecraft validation thresholds, and Wal\'en-style ratio cutoffs vary between studies and can change the same event's classification \citep{Hasegawa2024, Genestreti2025}. Any supervised model inherits whichever convention generated its labels.

The magnetosheath reconnection of interest here is the turbulence-driven type studied by \citet{Stawarz2019, Stawarz2022}: short, thin current sheets within an already turbulent flow rather than at a single organised boundary. This regime is significantly harder to search than magnetopause or magnetotail reconnection, which is preceded by minutes of organised inflow and produces ion-scale jets \citep{Paschmann1979, Yamada2010, Hesse2020}. Turbulence-driven crossings are sub-second, embedded in fluctuating background fields whose power spectrum spans the same scales as the crossing, and frequently of the electron-only type \citep{Phan2018, Hubbert2022} in which no ion jet is available. The multi-scale shape of the reversal is therefore the most reliable single-spacecraft feature, and the present detector is built around that shape rather than around an isolated peak in $|\vect{J}|$ or $|dB/dt|$.

Given the labeling disagreement noted above, we treat reconnection-event identification as a two-phase problem. \textit{Phase 1 (this paper)} compresses a large burst-mode search space down to a manageable set of physically plausible candidates. \textit{Phase 2 (subsequent work)} refines event boundaries and applies stronger physical and multi-spacecraft tests. The contribution here is a Phase-1 pipeline anchored to a single literature reference event and trained on a Monte Carlo library of physically calibrated surrogates of that event, offered here as an initial proof of concept before extending the one-class design to additional feature channels.

The remainder of the paper is organised as follows. Section~\ref{sec:data} introduces the MMS data and the 15 benchmark intervals; Section~\ref{sec:methods} develops the detection method; Section~\ref{sec:results} reports the operational outcome and example detections; Section~\ref{sec:discussion} discusses scope, limitations, and follow-up directions.

\section{Data and Benchmark Intervals}
\label{sec:data}

We use Magnetospheric Multiscale (MMS) burst-mode data from probe~1 \citep{Burch2016}, accessed through the open-source pySPEDAS package \citep{Grimes2022}. Three instruments are relevant. The Fluxgate Magnetometer (FGM) provides the magnetic field $\vect{B}$ at burst cadence \citep{Russell2016}. The Fast Plasma Investigation (FPI) provides ion and electron moments and bulk velocities \citep{Pollock2016} from which we compute the single-spacecraft current proxy
\begin{equation}
\vect{J} \;=\; e\,n\,(\vect{V}_{i} - \vect{V}_{e}),
\label{eq:Jproxy}
\end{equation}
where $e$ is the elementary charge, $n \equiv n_{i} \approx n_{e}$ is the quasineutral plasma number density, and $\vect{V}_{i}$ and $\vect{V}_{e}$ are the corresponding ion and electron bulk velocities. The Electric Double Probe (EDP) provides the electric field $\vect{E}$ when available \citep{Torbert2016}, from which we form the convective electric field
\begin{equation}
\vect{E}^{\prime} \;=\; \vect{E} + \vect{V}_{e}\times\vect{B},
\label{eq:Eprime}
\end{equation}
the electric field in the electron rest frame, and use the scalar product $\vect{J}\cdot\vect{E}^{\prime}$ as a diagnostic of energy conversion at the crossing.

We run the pipeline on 15 magnetosheath turbulence intervals listed in the supplementary table of \citet{Stawarz2022}, totalling \num{5671}~s (1.58~h) of burst-mode coverage between 2015 October and 2016 December (Table~\ref{tab:intervals}). The list is publicly catalogued and accompanied by an independent reconnection-event list against which the present pipeline can be cross-checked (Section~\ref{sec:results}); the cross-check measures how much of that curated multi-spacecraft list our single-spacecraft workflow recovers, not a head-to-head performance comparison.

The 15 intervals span Earth's dayside magnetosheath, from near the subsolar region (interval~4 at $X_{GSE}\!\approx\!72{,}000$~km, $Y_{GSE}\!\approx\!9{,}000$~km) to the dawn flank (intervals~6 and~7 at $X_{GSE}\!\approx\!12{,}000$--$13{,}000$~km, $Y_{GSE}\!\approx\!72{,}000$--$74{,}000$~km), coordinates given in the Geocentric Solar Ecliptic (GSE) system. Plasma conditions are typical of magnetosheath turbulence, with published ion plasma beta $\beta_{i}$ ranging from $\sim 1$ to $90$ across the set.

\begin{table*}[t]
\centering
\small
\caption{The 15 magnetosheath turbulence intervals adopted from \citet{Stawarz2022} and used as the scan target in this work. Durations are computed from the published start and end times. All scans use MMS1.}
\label{tab:intervals}
\begin{tabular}{cllrr}
\toprule
\# & Start time (UTC) & End time (UTC) & Duration (s) & Detections \\
\midrule
1  & 2015-10-21 06:55:10 & 2015-10-21 07:03:45 & 515 & 26 \\
2  & 2015-10-29 14:48:05 & 2015-10-29 14:53:55 & 350 & 4  \\
3  & 2015-10-31 10:22:27 & 2015-10-31 10:26:07 & 220 & 7  \\
4  & 2015-11-15 05:27:36 & 2015-11-15 05:32:10 & 274 & 19 \\
5  & 2016-02-23 21:54:06 & 2016-02-23 21:58:36 & 270 & 13 \\
6  & 2016-09-28 16:50:14 & 2016-09-28 17:00:14 & 600 & 23 \\
7  & 2016-09-30 17:50:34 & 2016-09-30 18:01:50 & 676 & 11 \\
8  & 2016-10-25 09:45:57 & 2016-10-25 09:50:07 & 250 & 3  \\
9  & 2016-11-09 13:41:23 & 2016-11-09 13:45:13 & 230 & 12 \\
10 & 2016-12-08 10:16:52 & 2016-12-08 10:21:02 & 250 & 12 \\
11 & 2016-12-08 10:25:02 & 2016-12-08 10:28:52 & 230 & 15 \\
12 & 2016-12-09 09:01:40 & 2016-12-09 09:07:00 & 320 & 24 \\
13 & 2016-12-09 09:26:25 & 2016-12-09 09:34:57 & 512 & 34 \\
14 & 2016-12-11 15:20:15 & 2016-12-11 15:32:31 & 736 & 44 \\
15 & 2016-12-31 13:27:04 & 2016-12-31 13:31:02 & 238 & 23 \\
\midrule
   & \multicolumn{2}{l}{\textbf{Total}} & \textbf{5671} & \textbf{270} \\
\bottomrule
\end{tabular}
\end{table*}

\section{Method}
\label{sec:methods}

\begin{figure}[p]
\centering
\includegraphics[height=0.82\textheight,width=\textwidth,keepaspectratio]{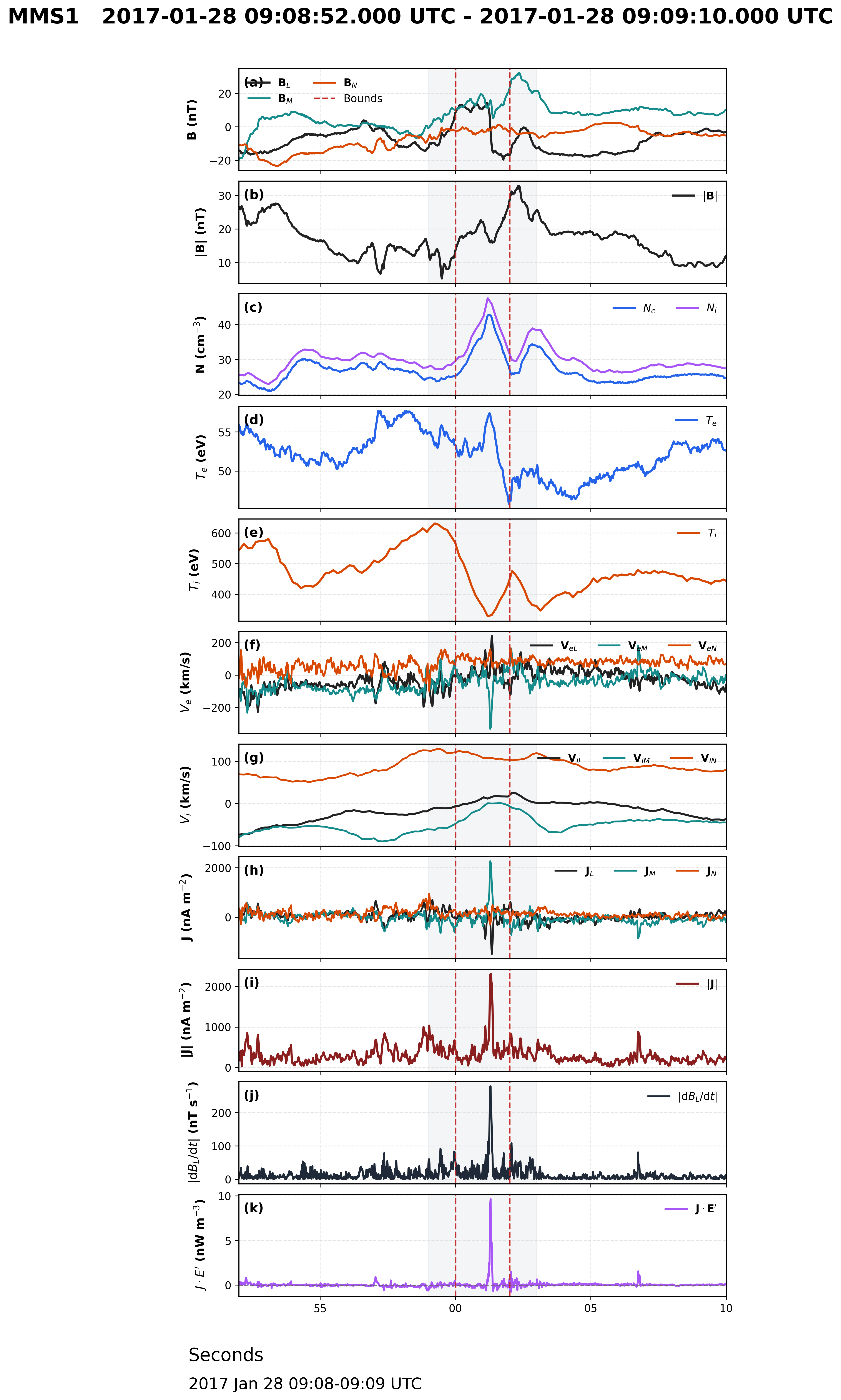}
\caption{Reference event (MMS1, 2017 Jan 28, 09:09:01~UTC) used to anchor the surrogate generator. Panels: (a) $\vect{B}$ in LMN coordinates and $|\vect{B}|$; (b) $n_{e}, n_{i}$; (c)--(d) $T_{e}, T_{i}$; (e)--(f) $\vect{V}_{e}, \vect{V}_{i}$ in LMN; (g) $\vect{J}$ from Equation~\ref{eq:Jproxy}; (h) $|\vect{J}|$; (i) $|dB_{L}/dt|$; (j) $\vect{J}\cdot\vect{E}^{\prime}$. Red dashed lines bound the 2~s analysis window. The $S_{2}(\tau)$ profile of the boxed window defines the acceptance band used by the surrogate generator (Equation~\ref{eq:s2band}).}
\label{fig:reference}
\end{figure}

\begin{figure*}[!htbp]
\centering
\includegraphics[width=\textwidth]{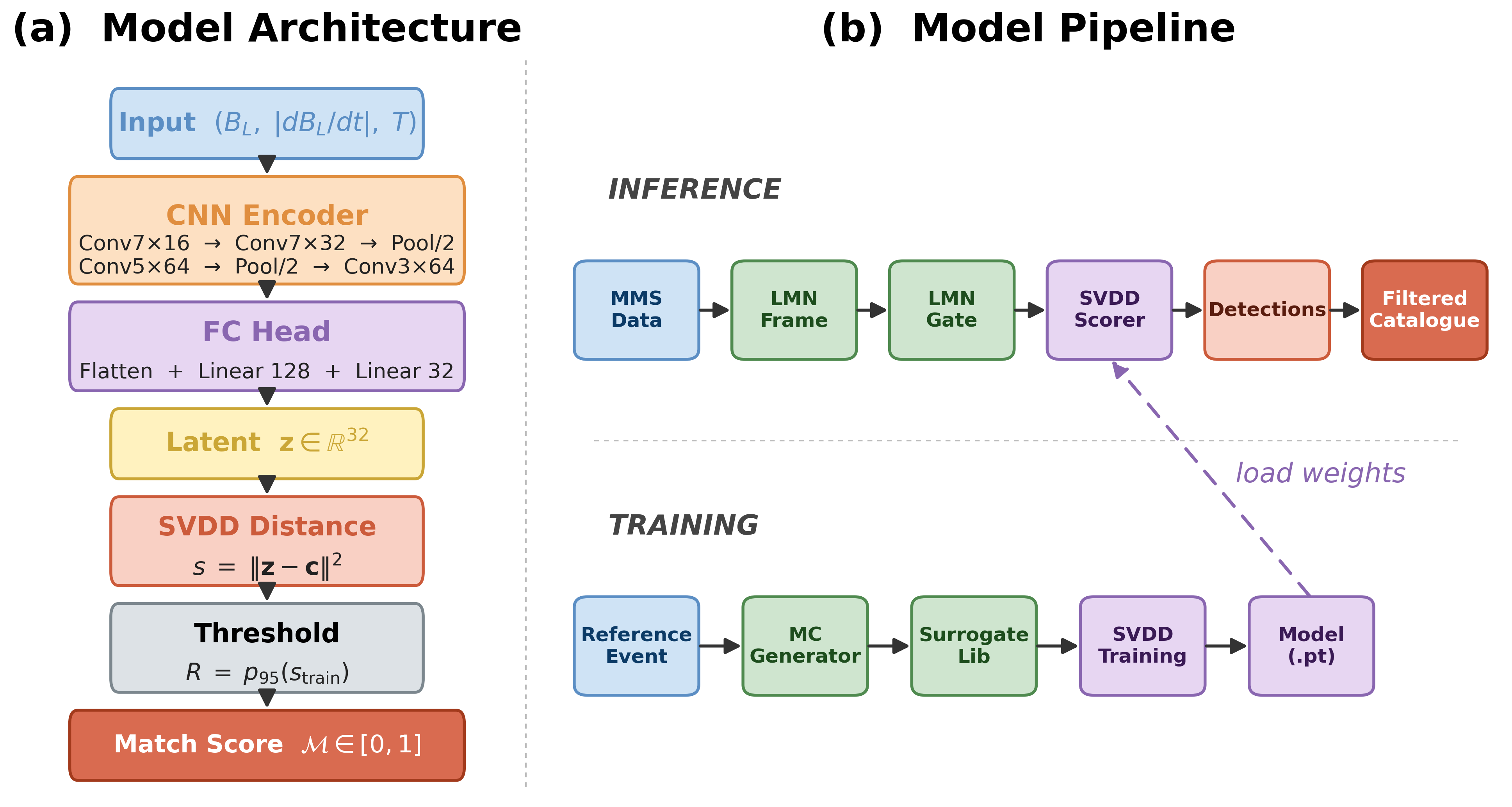}
\caption{End-to-end pipeline. Top row (inference): MMS burst-mode FGM, FPI, and EDP data are rotated into the local LMN frame by minimum-variance analysis (MVA), screened by an LMN-quality gate, scored by the trained SVDD model, grouped into packets, and reduced to one representative window per packet for downstream review. Bottom row (training): a single MMS reference event seeds a Monte Carlo surrogate generator; the resulting library of 3{,}000 calibrated $(B_{L},|dB_{L}/dt|)$ pairs is used to fit the one-class hypersphere whose checkpoint is loaded at inference (dashed arrow). The encoder is a small 1D convolutional network producing a 32-dimensional latent embedding $\vect{z}$; the SVDD distance $\|\vect{z}-\vect{c}\|^{2}$ is converted into a bounded match score in $[0,1]$ by Equation~\ref{eq:match}.}
\label{fig:svdd_arch}
\end{figure*}

The training distribution is anchored to one published MMS1 reconnection event, recorded on 28 January 2017 at 09:09:01~UTC and shown in Figure~\ref{fig:reference} in the same 11-panel layout used throughout the paper for individual detections. Figure~\ref{fig:svdd_arch} then summarises the end-to-end pipeline built around it, expressed in the LMN boundary-normal coordinate frame defined in Section~\ref{sec:method:seed} below. We describe each component in turn, working outward from this single reference event that anchors the training distribution.

\subsection{Reference event and the single-event design}
\label{sec:method:seed}

This event is the one shown in Figure~4 of \citet{Stawarz2022}. We work throughout in the local current-sheet frame (LMN coordinates), defined so that $\vect{L}$ is the magnetic-field reversal direction, $\vect{N}$ is the normal to the sheet, and $\vect{M} = \vect{N}\times\vect{L}$ closes the right-handed triad, obtained from minimum-variance analysis of the magnetic-field covariance over a 4~s window centered on the event, the same procedure used for the LMN-quality gate in Section~\ref{sec:method:lmn} \citep{Sonnerup1967, Sonnerup1998, Denton2018}. In this frame the reference event shows a sharp $B_{L}$ reversal with a co-located peak in $|\vect{J}|$ and an associated electron and ion flow response, consistent with the current-sheet morphology the pipeline is designed to find. The 2~s analysis window is bracketed by red dashed lines in Figure~\ref{fig:reference}. This event is well-suited as a seed because every signature the surrogate library needs is present in clean, isolated form: a sharp $B_{L}$ reversal that is well separated in time from neighbouring fluctuations, and a co-located $|\vect{J}|$ peak unambiguously associated with that reversal. The extensive diagnostic record already published for this event also provides an external check on every quantity the surrogate generator subsequently relies on.

A single reference event is a deliberate choice and warrants explanation. The detector ingests time-series \emph{shape}, not absolute amplitudes; what defines the in-class distribution is the multi-scale statistical fingerprint of the reversal, not the specific values of the electron bulk velocity $\vect{V}_{e}$ or the plasma density $n$ in the event. As long as that fingerprint is preserved, surrogate variation in amplitude, polarity, sheet thickness, and noise spectrum is exactly what we want the network to be invariant to. The Monte Carlo procedure described next is built around this idea: the generator produces a population whose internal variability spans the morphological diversity that real magnetosheath crossings show, while remaining anchored to a single physically clean template.

A previous version of this work embedded real events into a synthetic background and trained a supervised classifier on the contrast. That construction failed in two ways: the events and the background were not on the same statistical footing, so the network learned to discriminate the synthetic seam rather than sheet morphology and did not transfer cleanly to inference; and magnetosheath current sheets do not sit on a featureless background but on top of an underlying turbulent structure that a constant-background scaffold cannot reproduce. The present design avoids both failure modes by abandoning the separate-background construction: each surrogate is a single trace in which the magnetic reversal and the surrounding turbulent fluctuations come from the same construction, leaving no background-versus-event seam and no separate negative class to define.

\subsection{Monte Carlo surrogate library and $S_{2}(\tau)$ acceptance}
\label{sec:method:mc}

The generator builds each surrogate around a deterministic template motivated by the canonical asymmetric Harris current-sheet form \citep{Harris1962, CassakShay2007},
\begin{equation}
B_{L}(t) \;=\; A\,\tanh\!\left(\frac{t-t_{0}}{w}\right) + B_{\text{mid}} + \eta(t),
\label{eq:bl_template}
\end{equation}
where $A=(B_{2}-B_{1})/2$ is the half-jump across the sheet, $B_{\text{mid}}=(B_{1}+B_{2})/2$ is the midpoint of the unequal asymptotic field values appropriate to magnetosheath reconnection, $w$ is the sheet half-thickness, and $t_{0}$ is the crossing time. The amplitude, half-thickness, and crossing time are drawn from Gaussian posteriors centered on a least-squares fit to the reference event. To bring the synthetic morphology closer to that of observed magnetosheath sheets, the implementation also adds a small linear background drift and a small odd-polynomial term inside the $\tanh$ argument as bounded engineering modifications whose contribution lies below the level of the noise variability. The noise $\eta(t)$ is a two-band coloured-noise process modulated by a core-suppression envelope: the two bands reproduce the inertial-range and kinetic-range slopes of the seed event's $S_{2}(\tau)$ (a low-frequency band with shallower slope mimics large-scale fluctuations in the inflow regions, a high-frequency band with steeper slope mimics kinetic-scale fluctuations on which electron physics develops), and the envelope multiplies the noise by a smooth window centred on the reversal that approaches zero inside the core, so the final surrogate has a clean reversal embedded in a statistically calibrated turbulent background rather than a noise-corrupted reversal. The candidate is bandlimited and given a random sign; the second training channel is the absolute derivative $|dB_{L}/dt|$, which serves as a current-density proxy at fixed cadence and is sign-invariant.

The acceptance test for any candidate is its second-order structure function,
\begin{equation}
S_{2}(\tau) \;=\; \big\langle \big[B_{L}(t+\tau)-B_{L}(t)\big]^{2}\big\rangle_{t},
\label{eq:s2def}
\end{equation}
a multi-scale measure of how strongly the field changes across a time lag $\tau$ \citep{Frisch1995, Kolmogorov1941, Alexandrova2012}. $S_{2}(\tau)$ acts as a multi-scale fingerprint: a current-sheet crossing imprints a characteristic pattern in $\log S_{2}(\tau)$ versus $\log\tau$, with a steep rise at the small-lag scales that resolve the reversal core and a flatter slope at larger lags that average over the inflow regions. Two events that share this multi-scale shape are statistically alike under $S_{2}(\tau)$ even when they differ in amplitude, polarity, or precise timing, and that property is what the surrogate generator exploits. We make this amplitude invariance explicit: the reference event is amplitude-normalised before $S_{2}(\tau)$ is computed, every surrogate is normalised on the same footing before the acceptance test is applied, and each candidate window at inference time is rescaled by the same procedure before it is scored, so the network never sees absolute field magnitudes. Higher-order generalisations of $S_{2}(\tau)$ have a long history of use in characterising kinetic-scale plasma turbulence in spacecraft observations \citep{Uritsky2011}, and bulk turbulence-statistic measures have been used in adjacent contexts to discriminate plasma structures, for example to separate solar-wind background, coronal mass ejection (CME) sheaths, and magnetic clouds via Fourier power \citep{Ghuge2025}. A candidate is accepted only if its $S_{2}(\tau)$ stays inside an acceptance band around the reference,
\begin{equation}
\big|\log_{10} S_{2}^{\text{cand}}(\tau) - \log_{10} S_{2}^{\text{ref}}(\tau)\big| \;\le\; 0.20\;\text{dex}\quad\text{for at least 60\% of lags},
\label{eq:s2band}
\end{equation}
together with five shape-based checks. (1) A ramp/curvature check rejects candidates that are too close to a straight line (a first-order fit to $B_{L}(t)$ explaining more than 99.8\% of its variance) or too weakly curved (the ratio of the mean absolute second derivative to the mean absolute first derivative below $2\times10^{-3}$), since a genuine reversal has the S-shaped bend of a current sheet rather than a linear trend. (2) A multi-peak veto counts local maxima of $|dB_{L}/dt|$ exceeding 35\% of the trace's peak gradient and rejects candidates with more than three such peaks, keeping the library to single, isolated reversals rather than multi-crossing traces. (3) A kinetic-range slope match requires the log-log slope of the candidate's $S_{2}(\tau)$ over lags of 2--70 samples to lie within 0.6 of the same slope computed from the reference event, applied jointly with the amplitude-band criterion of Equation~\ref{eq:s2band}, so that the small-scale statistics of every surrogate track the multi-scale fingerprint the detector is trained to recognise. (4) A clone check computes the sign-invariant mean-squared error and the Pearson correlation between the candidate and the reference trace, rejecting near-duplicates (mean-squared error below 0.02, or absolute correlation above 0.985) that would otherwise let the sampler return trivial near-copies of the single seed event instead of adding morphological diversity. (5) A high-frequency leakage check low-pass filters the candidate at 5~Hz, defines the reversal core as the samples where $|dB_{L}/dt|$ exceeds 55\% of its peak, and rejects candidates whose high-frequency residual is more than 35\% as strong (in RMS terms) inside the core as on the flanks; this keeps the reversal itself clean even though noise is deliberately injected around it to mimic turbulence. As with the tolerance values above, the specific thresholds used in these five checks are engineering choices rather than the product of a systematic sensitivity study: they were tuned so that visually acceptable candidates passed and visually implausible ones were rejected, and could reasonably be set differently without changing the qualitative behaviour of the generator. The numeric tolerance values $\{0.20\;\text{dex},\,60\%,\,\pm 0.6\}$ are similarly engineering choices: the dex band is wide enough to cover physically reasonable variability in the inertial range while still excluding profiles whose multi-scale energy budget is structurally different from the seed, and the 60\% lag-coverage rule allows narrow excursions outside the band without admitting candidates that wander out at multiple scales. We do not claim that $S_{2}(\tau)$ alone uniquely identifies sheet-like morphology: in principle, a different time-domain profile could produce a similar structure function. The five shape-based checks above, and, at inference time, the SVDD encoder's use of the full two-channel time series rather than $S_{2}(\tau)$ alone (Section~\ref{sec:method:svdd}), provide additional, largely independent discrimination; the residual risk of a profile that passes both is caught downstream by the mandatory visual screening described in Section~\ref{sec:results:visual}.

\begin{figure*}[!htbp]
\centering
\includegraphics[width=\textwidth]{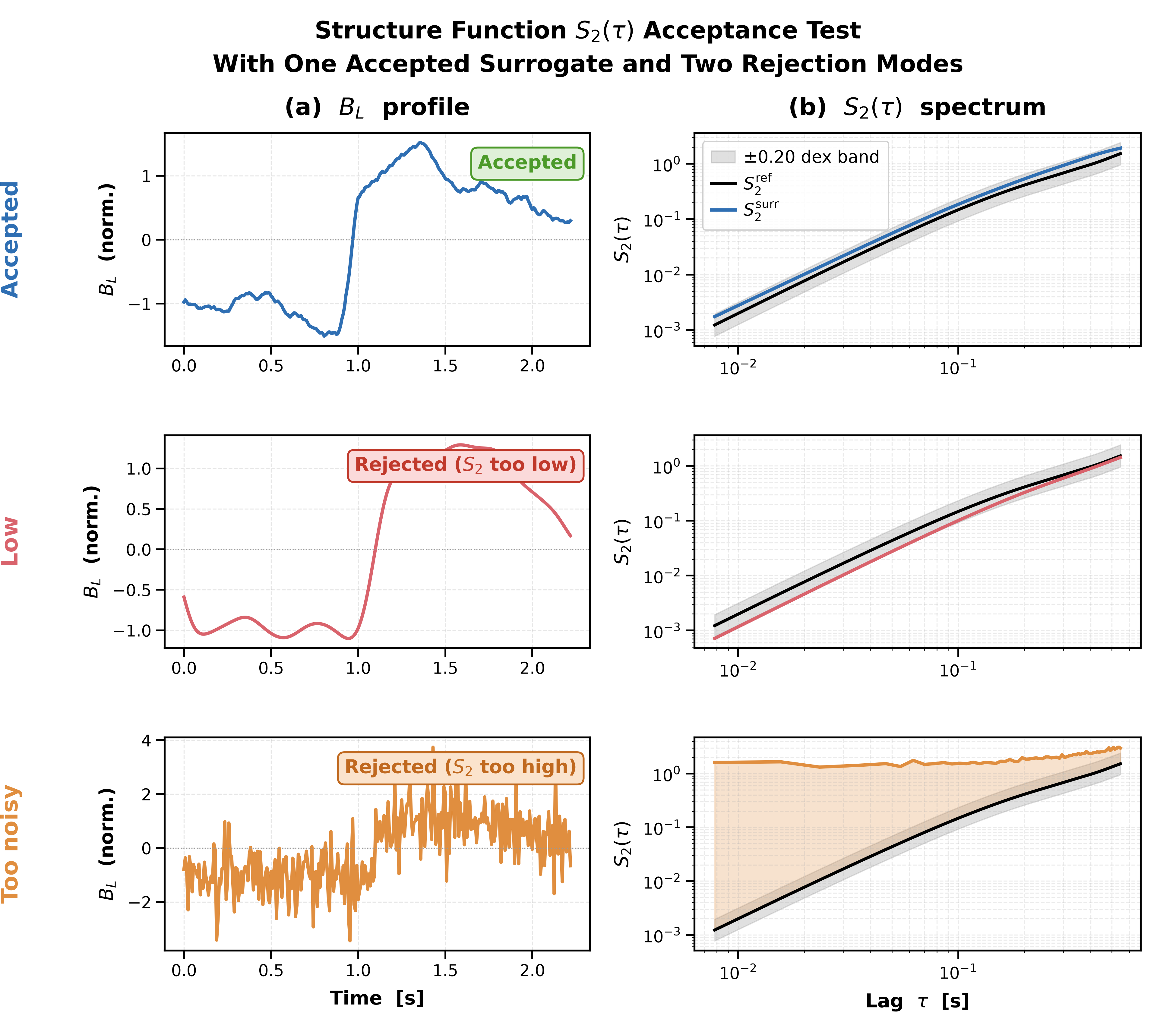}
\caption{Structure-function acceptance test. (a) $B_{L}(t)$ profile for an accepted surrogate (top, blue), a low-$S_{2}$ candidate that strips small-scale variability (middle, salmon), and an over-noisy candidate that lets broadband fluctuations leak into the reversal core (bottom, orange). (b) Corresponding $S_{2}(\tau)$ on log-log axes. The grey band is the $\pm 0.20$~dex tolerance around the reference $S_{2}(\tau)$ (black). The accepted surrogate tracks the reference at every lag; the low-$S_{2}$ candidate falls below the band at small lags; the over-noisy candidate rises above it. Shaded violations make each failure mode visually explicit.}
\label{fig:s2_explainer}
\end{figure*}

Figure~\ref{fig:s2_explainer} illustrates the acceptance test on three example traces: one candidate accepted into the surrogate library, together with two constructed illustrations of the rejection modes rather than generator outputs themselves: a low-$S_{2}$ trace (which falls below the band at small lags because removing small-scale variability lowers the local mean-squared difference) and an over-noisy trace (which rises above the band at small lags because broadband fluctuations inflate the small-lag squared differences). The lower edge of the band matters for a physical reason, not just a statistical one: loosening it to admit smoother candidates would mean training on profiles that lack the finite small-scale roughness real current-sheet crossings retain at the reversal, pushing the accepted population toward an idealised, step-like tanh shape rather than the somewhat-rough transitions actually observed, even in comparatively quiescent turbulence.

\begin{figure*}[!htbp]
\centering
\includegraphics[width=\textwidth]{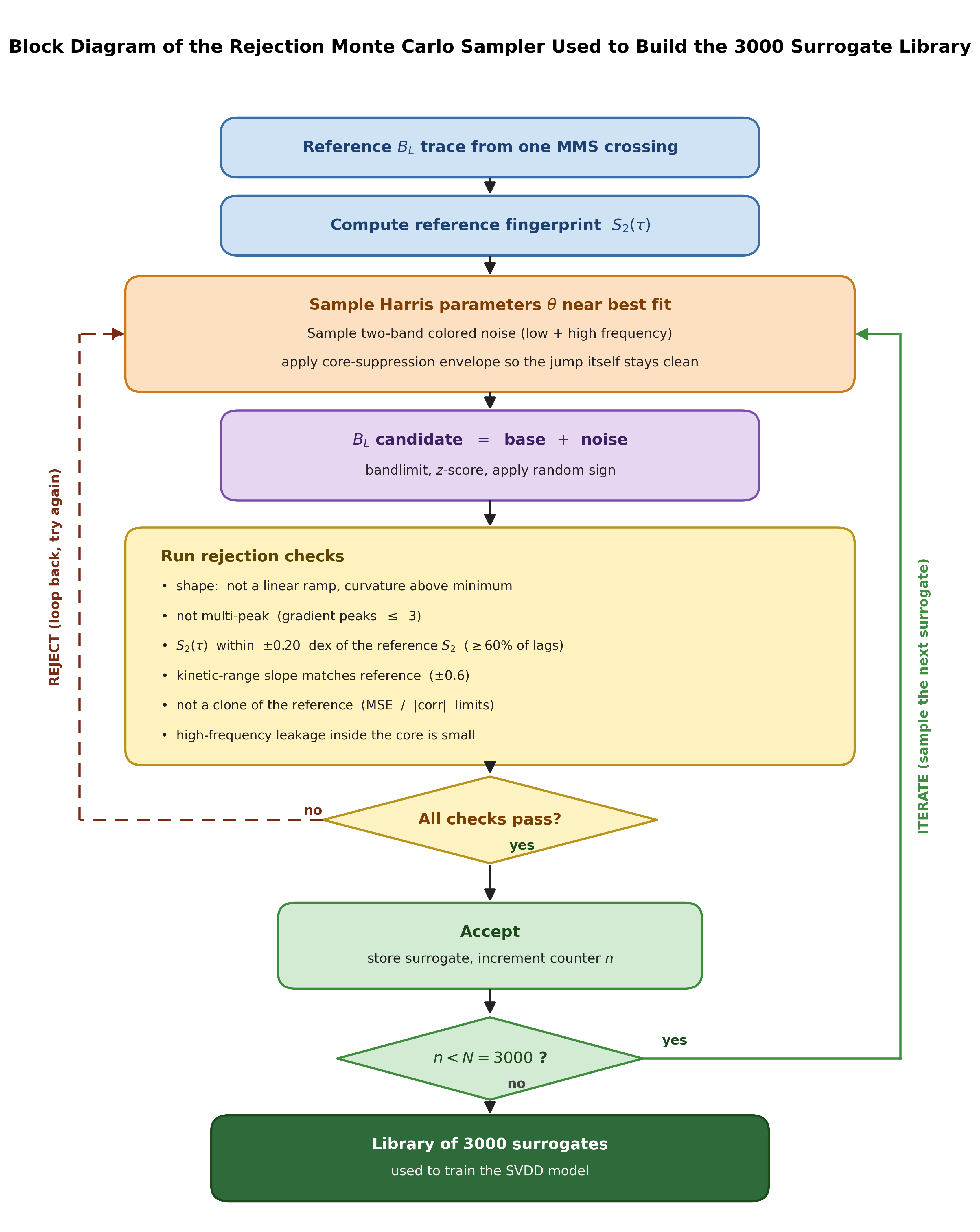}
\caption{Block diagram of the rejection-sampling Monte Carlo generator. Inside the sampling loop (left), Harris parameters and a two-band coloured-noise process produce a candidate $B_{L}$ trace, which is bandlimited, $z$-scored, and given a random sign. Each candidate is then evaluated against six rejection criteria (centre): a shape check, a multi-peak veto, the spectral acceptance test (Equation~\ref{eq:s2band}), a kinetic-range slope match, a clone check, and a low high-frequency leakage check inside the reversal core. Surrogates that pass all six are added to the library (right); the loop runs until $N=3000$ accepted samples are stored.}
\label{fig:mc_block}
\end{figure*}

The full sampling loop is shown in Figure~\ref{fig:mc_block}. Inside the loop, Harris parameters and a two-band coloured-noise process produce a candidate $B_{L}$ trace which is then evaluated against six rejection criteria; surrogates that pass all six are added to the library, and the loop continues until $N=3000$ surrogates have been accepted. The library size is chosen so that the sorted training-distance distribution used to compute the match score's percentile rank is statistically stable and so that the morphology coverage of the surrogate distribution is empirically saturated; pushing $N$ above $10^{4}$ thickens this distribution without widening its support and offers diminishing returns. Acceptance fractions are typically a few percent, so a 3{,}000-event library requires of order $10^{5}$ trial draws.

\begin{figure*}[!htbp]
\centering
\includegraphics[width=\textwidth]{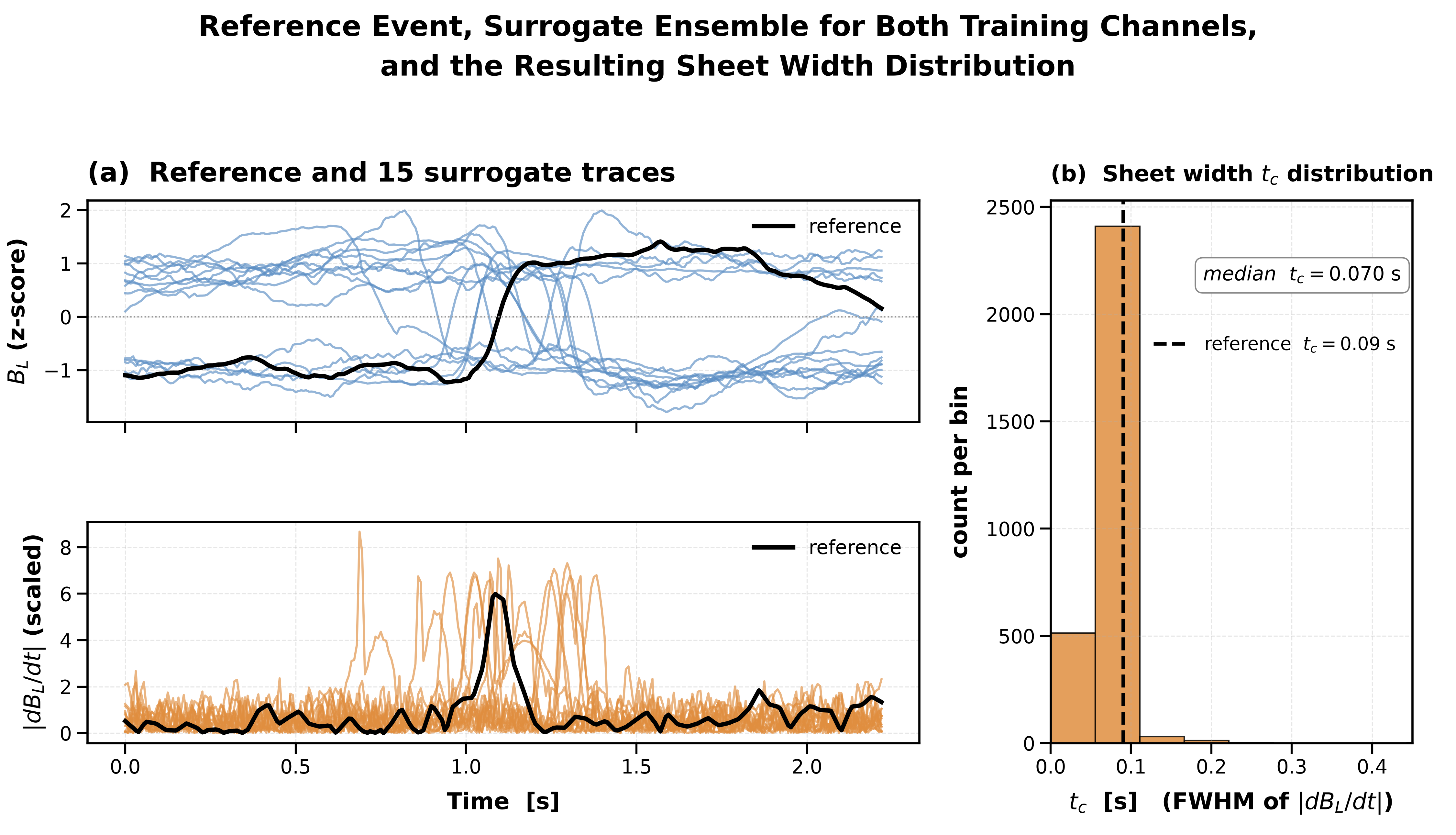}
\caption{Surrogate ensemble. (a) Reference trace (black) and 15 surrogate traces for both training channels: $B_{L}$ (top) and $|dB_{L}/dt|$ (bottom). The surrogates show clear shape diversity around the reference while preserving the multi-scale fingerprint by construction. (b) FWHM-based sheet-width distribution across all 3{,}000 generated surrogates; the dashed line marks the reference value. The ensemble is sharply peaked around the reference, confirming that the rejection sampler preserves the characteristic sheet thickness of the seed event.}
\label{fig:surrogate_ensemble}
\end{figure*}

The resulting library is the only training set the SVDD model ever sees. Figure~\ref{fig:surrogate_ensemble} shows the diversity it captures: panel~(a) overlays 15 surrogate traces on the reference for both training channels, and panel~(b) shows the full width at half maximum (FWHM)-based sheet-width distribution of all 3{,}000 surrogates, sharply peaked near the reference value $w\!\approx\!0.07$~s.

\subsection{One-class scoring with Deep SVDD}
\label{sec:method:svdd}

Because there is no curated, statistically representative ``not-a-current-sheet'' negative class in turbulent magnetosheath data, we adopt \emph{one-class} anomaly detection: a method that learns only from positive (sheet-like) examples and scores how similar new inputs are to those examples. The picture is geometric. A learned function $\phi_{\theta}$ maps each 2~s window to a single point in a 32-dimensional latent space, and the training procedure adjusts $\phi_{\theta}$ so that all 3{,}000 surrogate points collapse into a tight ball around a fixed centre $\vect{c}$. At inference, an arbitrary new window is passed through the same $\phi_{\theta}$, and its squared distance from $\vect{c}$, $\|\phi_{\theta}(\vect{x})-\vect{c}\|^{2}$, becomes a one-number measure of similarity: windows that resemble the trained surrogates land close to $\vect{c}$, dissimilar windows land further out. This formulation is the Deep Support Vector Data Description (Deep SVDD) framework \citep{TaxDuin2004, Scholkopf2001, Ruff2018}, and we adopt it for the geometric interpretability of its score \citep{Pang2022}.

The encoder $\phi_{\theta}$ is a small 1D convolutional network (a stack of learned filters that slide along the time series and progressively compress its shape into a fixed-length vector) operating on the two-channel input $(B_{L}, |dB_{L}/dt|)$ at 128~Hz over $T\!=\!285$ samples (the $\sim$2~s scan window). Its output is a 32-dimensional latent embedding $\vect{z} = \phi_{\theta}(\vect{x})\in\mathbb{R}^{32}$. Training adjusts the network parameters $\theta$ to minimise the joint loss
\begin{equation}
\mathcal{L}(\theta)
\;=\;
\underbrace{\frac{1}{N}\sum_{i=1}^{N}\bigl\|\phi_{\theta}(\vect{x}_{i})-\vect{c}\bigr\|^{2}}_{\text{SVDD pull}}
\;+\;
\underbrace{\lambda\,\mathcal{L}_{\text{recon}}(\theta)}_{\text{regulariser}},
\quad \lambda=0.25,
\label{eq:svdd_loss}
\end{equation}
where the first term (the ``SVDD pull'') contracts the cloud of training points towards $\vect{c}$, and $\mathcal{L}_{\text{recon}}$ is a small reconstruction penalty that requires the network to be able to rebuild its input from the latent point. This regulariser blocks the trivial-collapse failure mode in which the encoder would minimise the SVDD pull artificially by mapping every input to $\vect{c}$. Training proceeds in two phases: 8 epochs of reconstruction-only pre-training to give the latent space a sensible geometry, followed by 30 epochs of the joint loss using the AdamW optimiser \citep{Loshchilov2019}. The decision threshold is set internally from the training data itself: at the end of training, we compute the distance of every surrogate to $\vect{c}$ and sort these into $\{s_{1}\le s_{2}\le\dots\le s_{N}\}$. At inference, an arbitrary 2~s burst-mode window $\vect{x}$ is passed through the same encoder, and its distance $d(\vect{x}) = \|\phi_{\theta}(\vect{x})-\vect{c}\|^{2}$ is converted into its empirical percentile rank within the training distribution,
\begin{equation}
P(\vect{x}) \;=\; 100 \times \frac{\#\{s_{i}\le d(\vect{x})\}}{N},
\label{eq:pct}
\end{equation}
which is in turn converted into a bounded match score
\begin{equation}
\mathcal{M}(\vect{x})
\;=\;
1 - \frac{P(\vect{x})}{100}\;\in\;[0,1],
\label{eq:match}
\end{equation}
so that a window whose distance ranks below all of the training surrogates scores $\mathcal{M}\approx 1$, a window whose distance sits at the 95th percentile of the training cloud scores $\mathcal{M}\approx 0.05$, and a window at or beyond the single most distant training surrogate scores $\mathcal{M}=0$. A window is reported as a positive detection if $\mathcal{M}(\vect{x})\ge 0.7$, i.e.\ its distance ranks below the 30th percentile of the training cloud.

The value 0.7 was chosen pragmatically: we evaluated thresholds of 0.3, 0.5, 0.7, and 0.85 and observed qualitatively that lower thresholds admitted many low-quality detections that were costly to review, while 0.85 excluded a non-trivial fraction of clear sheet-like crossings. A systematic sensitivity analysis of this threshold is left for future work.

\subsection{LMN-quality gate, sliding scan, and grouping}
\label{sec:method:lmn}

The SVDD score is informative only when the LMN frame in which $B_{L}$ is computed is itself well defined. If the minimum-variance eigenvalues are not well separated, the choice of L direction is ambiguous, $B_{L}$ becomes a noisy combination of components rather than the clean reversal direction the SVDD scorer was trained on, and the output score loses its physical meaning. We therefore apply a frame-quality preselector before scoring. For each candidate 2~s window centred at midpoint $t_{c}$, we form a slightly broader 4~s minimum-variance interval $[t_{c}-2,\,t_{c}+2]$~s and compute its magnetic-field covariance \citep{Sonnerup1967, Sonnerup1998, Denton2018}. Sorting the eigenvalues $\lambda_{1}\ge\lambda_{2}\ge\lambda_{3}$, we form the ratios
\begin{equation}
r_{12} \;=\; \lambda_{1}/\lambda_{2}, \qquad r_{23} \;=\; \lambda_{2}/\lambda_{3},
\label{eq:rratios}
\end{equation}
and declare the window LMN-pass if $r_{12}\ge 5$ and $r_{23}\ge 5$, chosen so that the local L direction and the normal direction $\vect{N}$ are each well separated from the others. Windows that fail this gate are not scored. Crucially, the gate does not require a hard $B_{L}=0$ crossing or any explicit reconnection criterion at this stage; it is a frame-quality prefilter, not a final reconnection test.

\begin{figure*}[!htbp]
\centering
\includegraphics[width=\textwidth]{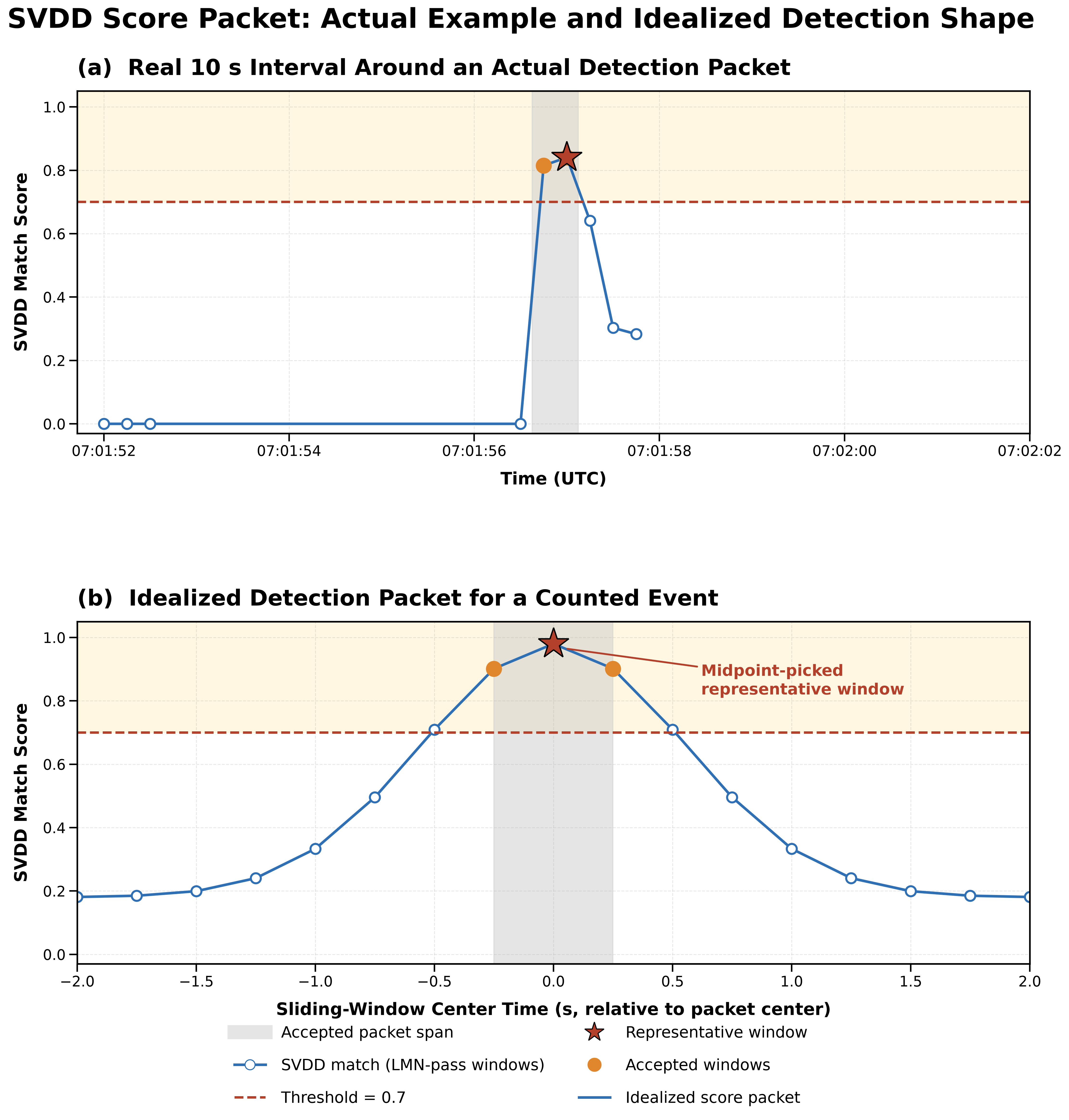}
\caption{Score-packet logic. (a) A 10~s sliding-window scan covering 2015-10-21 07:01:52.752--07:02:02.752~UTC and centred on a real detected packet. Blue circles show SVDD match scores for LMN-pass windows; the dashed red line is the threshold ($\mathcal{M}\ge 0.7$); orange markers are accepted windows; the red star is the representative midpoint window retained as the final detection. (b) Idealised schematic of the same logic: a locally elevated score packet rises above threshold for several adjacent strides, and the representative detection is the accepted window nearest the temporal midpoint.}
\label{fig:score_packet}
\end{figure*}

The scan is at fixed cadence with a 2~s window and a 0.25~s stride. The reversal core itself is shorter (the FWHM of $|dB_{L}/dt|$ in the seed event is approximately 0.07~s), but a window short enough to be local while still containing both inflow shoulders needs of order one second on each side of the reversal so that the shoulder regions used by the structure-function acceptance test of Section~\ref{sec:method:mc} are statistically meaningful. A narrower window would drop the multi-scale fingerprint that the surrogate generator preserves; a broader one would average over neighbouring crossings in the turbulent flow and wash out the morphology. LMN-pass windows whose match score satisfies Equation~\ref{eq:match} above 0.7 are merged into packets whenever their pass-window centres lie within a merge gap of 0.3~s. Within each packet, the representative window stored as the final detection is the one whose centre lies closest to the packet midpoint
\begin{equation}
t_{\text{mid}} \;=\; \tfrac{1}{2}\!\left[\min(t_{c,i}) + \max(t_{c,i})\right].
\end{equation}
This logic is illustrated in Figure~\ref{fig:score_packet}: a real burst-survey packet from interval~1 with two pass windows above threshold, and an idealised schematic of the same midpoint-selection rule.

\subsection{Manual workup of detections}
\label{sec:method:manual}

For each retained detection the pipeline saves an 11-panel plot, a per-detection time-series CSV, and a metadata JSON sufficient to reproduce the window. We then inspect each plot and assign one of three labels using a deliberately simple visual criterion (Section~\ref{sec:results:visual}). For a small number of priority detections we additionally re-run a standalone single-spacecraft workup that the user invokes by manually placing red dashed boundaries inside the saved plot. That workup recomputes local plasma scales $d_{e}$ and $d_{i}$ from the local densities, ion and electron jet differences across the manually selected boundaries, the Cassak--Shay asymmetric inflow Alfv\'en speed \citep{CassakShay2007}
\begin{equation}
V_{A,\text{inflow}}
\;=\;
\sqrt{\frac{|B_{L,1}||B_{L,2}|\,(|B_{L,1}|+|B_{L,2}|)}
{\mu_{0}\,m_{i}\,(n_{1}|B_{L,2}|+n_{2}|B_{L,1}|)}},
\label{eq:vainflow}
\end{equation}
where $B_{L,1}$ and $B_{L,2}$ are the L-component magnetic fields measured on the two sides of the chosen interval, $n_{1}$ and $n_{2}$ are the corresponding ion number densities, $\mu_{0}$ is the vacuum magnetic permeability, and $m_{i}$ is the proton mass; and a single-spacecraft proxy of the Wal\'en jet ratio $|\Delta u_{eL}|/|\Delta V_{A,L}|$, where $\Delta u_{eL}$ is the larger of the two deviations between the electron-flow speed at the $B_{L}=0$ crossing and its value averaged over each adjacent shoulder, and $\Delta V_{A,L}$ is the analogous shoulder-to-crossing deviation of the local Alfv\'en speed profile $V_{A,L}(t) = B_{L}(t)/\sqrt{\mu_{0} m_{i} n(t)}$. The physical content of this ratio is that an actively reconnecting current sheet accelerates plasma to a fraction of the local Alfv\'en speed, so a jet ratio close to unity is the signature one expects from an active reconnection event under the simplest single-spacecraft assumptions. The simple decision rule we apply on the manually selected interval has two parts: the magnetic field must reverse, $B_{L,1}\,B_{L,2}<0$, and the electron jet ratio must satisfy $|\Delta u_{eL}|/|\Delta V_{A,L}|\ge 0.7$. Both of these are simplified single-spacecraft adaptations of criteria used in earlier four-spacecraft analyses of magnetosheath reconnection \citep{Stawarz2022}, and we emphasise that this advisory check is not a substitute for that four-spacecraft analysis. We use it only to characterise individual example events and to confirm that the simple visual labels are physically consistent with width-and-jet diagnostics on the chosen boundaries.

\section{Results}
\label{sec:results}

\subsection{Search-space reduction}
\label{sec:results:reduction}

Across the 15 turbulence intervals (Table~\ref{tab:intervals}), the fixed 2~s / 0.25~s scan visited \num{22775} sliding windows. After the LMN-quality gate, SVDD scoring at $\mathcal{M}\ge 0.7$, and packet grouping with merge gap 0.3~s, the pipeline returned 270 final detections, a total compression of 98.8\%. Per-interval detection counts are reported in the rightmost column of Table~\ref{tab:intervals} and range from 3 in interval~8 (250~s of magnetosheath) to 44 in interval~14 (736~s).

It is informative to break the total compression down into its three stages, because the LMN-quality gate, the SVDD score, and the grouping rule each contribute differently. Across the 15 intervals the LMN gate alone passes 3{,}103 of the \num{22775} scanned windows (13.6\%); applying the SVDD threshold $\mathcal{M}\ge 0.7$ retains 773 of those LMN-pass windows (24.9\% of the LMN-pass population, or 3.4\% of the scan); and packet grouping reduces the SVDD-pass population to the 270 retained detections (1.2\% of the scan). The LMN-quality gate therefore performs the largest single compression in the chain, the SVDD score provides an additional factor of about four on the LMN-pass population, and grouping a further factor of about three. The SVDD score is not a redundant check on the LMN gate: across the 15 intervals it rejects between 25\% and 96\% of LMN-pass windows depending on interval, with a median rejection rate of 75\%.

\subsection{Visual screening criterion and counts}
\label{sec:results:visual}

We labeled all 270 detections by inspecting the 11-panel plot of each detection. The criterion is intentionally simple, and applied in the order shown.

A detection is assigned label~2, sheet-like, if both (i) $B_{L}$ shows a clear sharp change across the marked window, often reversing sign or rotating strongly, and (ii) $|\vect{J}|$ shows a clear local enhancement near that same window. A detection is assigned label~1, candidate reconnection, if it satisfies the sheet-like criteria above and additionally shows (iii) a clear localised electron-flow change or jet near the same window, (iv) magnetic change, current enhancement, and flow signature that line up to a visible extent, and (v) $\vect{J}\cdot\vect{E}^{\prime}>0$ inside the marked window. A detection is assigned label~3, not convincing, when $B_{L}$ does not show a clear sharp change across the window, or $|\vect{J}|$ is not locally enhanced, or the flow signature is absent, or the feature is too narrow or too broad to be interpreted with confidence.

The breakdown is summarised in Table~\ref{tab:results_summary}. Of the 270 detections, 93 were assigned label~1 (candidate reconnection), 118 label~2 (sheet-like), and 59 label~3 (not convincing). The combined sheet-or-better population (labels 1+2) accounts for 78.1\% of the reviewed set, while label~3 accounts for 21.9\%. Of the original \num{22775} scanned windows, the candidate-reconnection subset retained after manual review represents 0.41\%, a useful order-of-magnitude estimate of how rare the strict ``visually reconnection-like'' morphology is in this dataset.

\begin{table}[t]
\centering
\small
\caption{Compact summary of the 15-interval scan.}
\label{tab:results_summary}
\begin{tabular}{lr}
\toprule
Quantity & Value \\
\midrule
Analysed intervals & 15 \\
Total burst duration & 5{,}671~s (1.58~h) \\
Sliding windows evaluated & \num{22775} \\
Final detections after grouping & 270 \\
Search-space reduction & 98.8\% \\
Candidate reconnection (label 1) & 93 (34.4\%) \\
Sheet-like (label 2) & 118 (43.7\%) \\
Sheet-or-better (labels 1+2) & 211 (78.1\%) \\
Not convincing (label 3) & 59 (21.9\%) \\
Catalog events touched & 24 (out of overlapped subset) \\
Detections overlapping a catalog event & 22 \\
Visual recall on the overlap subset & 81.8\% \\
\bottomrule
\end{tabular}
\end{table}

\subsection{What the SVDD score appears to measure}
\label{sec:results:scores}

A useful consistency check is whether the SVDD score itself separates labels 1 and 2 within the manually retained pool. It does not. The mean match score among label-1 detections was 0.913, and among label-2 detections it was 0.922. The two distributions overlap almost completely. The interpretation is straightforward and consistent with the design: once the LMN-quality gate has restricted the search to plausible local current-sheet frames, the SVDD score discriminates sheet-like \emph{morphology} but cannot, by itself, separate reconnecting from non-reconnecting sheets. A current sheet is necessary but not sufficient for reconnection, and a single-spacecraft morphology scorer is not the right tool for that distinction. The label-1-vs-label-2 separation is left to the visual flow-and-energy criterion described above.

\subsection{Cross-check against a known published catalog}
\label{sec:results:catalog}

The 93 candidate-reconnection detections of Section~\ref{sec:results:visual} are the principal output of the pipeline, and the natural next question is how they relate to an independent reconnection-event catalog over the same intervals. We use the supplementary reconnection-event list of \citet{Stawarz2022} as such a sanity check. That list was constructed by scanning 60 magnetosheath turbulence intervals across all four MMS spacecraft in a two-step process: an automated search retained a candidate only when the magnetic-field reversal and the associated local current peak were observed consistently on all four spacecraft and the measured electron jet across the crossing was compatible with the asymmetric-reconnection prediction, after which each retained candidate was manually examined for subtler signatures of reconnection, including the consistency of Hall-field perturbations, deflection of the electron jet by the Lorentz force associated with the guide field, and energy-conversion signatures; the published list contains 256 reconnection events, of which 18 had a coupled ion jet. The protocol is therefore systematically more conservative than the present single-spacecraft scan, so it is not the same target the present method aims at, and partial overlap is the expected outcome rather than a yield estimate. Of our 270 detections, 22 detection windows overlap one or more MMS1 reconnection events in the published list, collectively touching 24 unique catalog rows. Of the published catalog's 78 events that fall within these same 15 intervals, this is a recall of 24/78 (30.8\%). This gap is expected given the design choices described above: the LMN-quality gate alone passes only 13.6\% of scanned windows (Section~\ref{sec:results:reduction}), so a catalog event whose local field geometry does not happen to produce a well-separated minimum-variance frame within our fixed 2~s cadence is excluded before it is ever scored, independent of whether it is a genuine reconnection event; the one-class scorer, anchored to a single reference event's morphology, may also underweight catalog events whose sheet thickness, cleanliness, or noise character differs from the training population. Eighteen of those 22 overlap detections were also assigned label~1 in our blind manual screening, which gives 81.8\% agreement on the overlap subset.

The headline reading of these numbers is that the candidate-reconnection pool the present pipeline surfaces is substantially larger than what any single curated catalog records over the same intervals: 93 candidate detections in our list, of which 18 also appear in the published catalog. The remaining 75 candidate-reconnection detections did not overlap published events. This is consistent with the design of the two methods. The published catalog uses a different event-finding logic and a four-spacecraft validation step, which is more conservative by construction; our pipeline is single-spacecraft and is deliberately tuned to surface a richer set of candidates for human review. A subset disagreement between the two outputs is therefore expected, and the additional candidates the present method surfaces are exactly the set that would benefit from follow-up multi-spacecraft analysis. Specific pipeline choices that contribute to the catalog-overlap rate (the fixed 2~s scan window, the strict frame-quality gate $r_{12},r_{23}\ge 5$, and the absence of an automatic boundary-refinement step) are discussed in Section~\ref{sec:discussion}.

\subsection{Examples}
\label{sec:results:examples}

\begin{figure}[p]
\centering
\includegraphics[height=0.82\textheight,width=\textwidth,keepaspectratio]{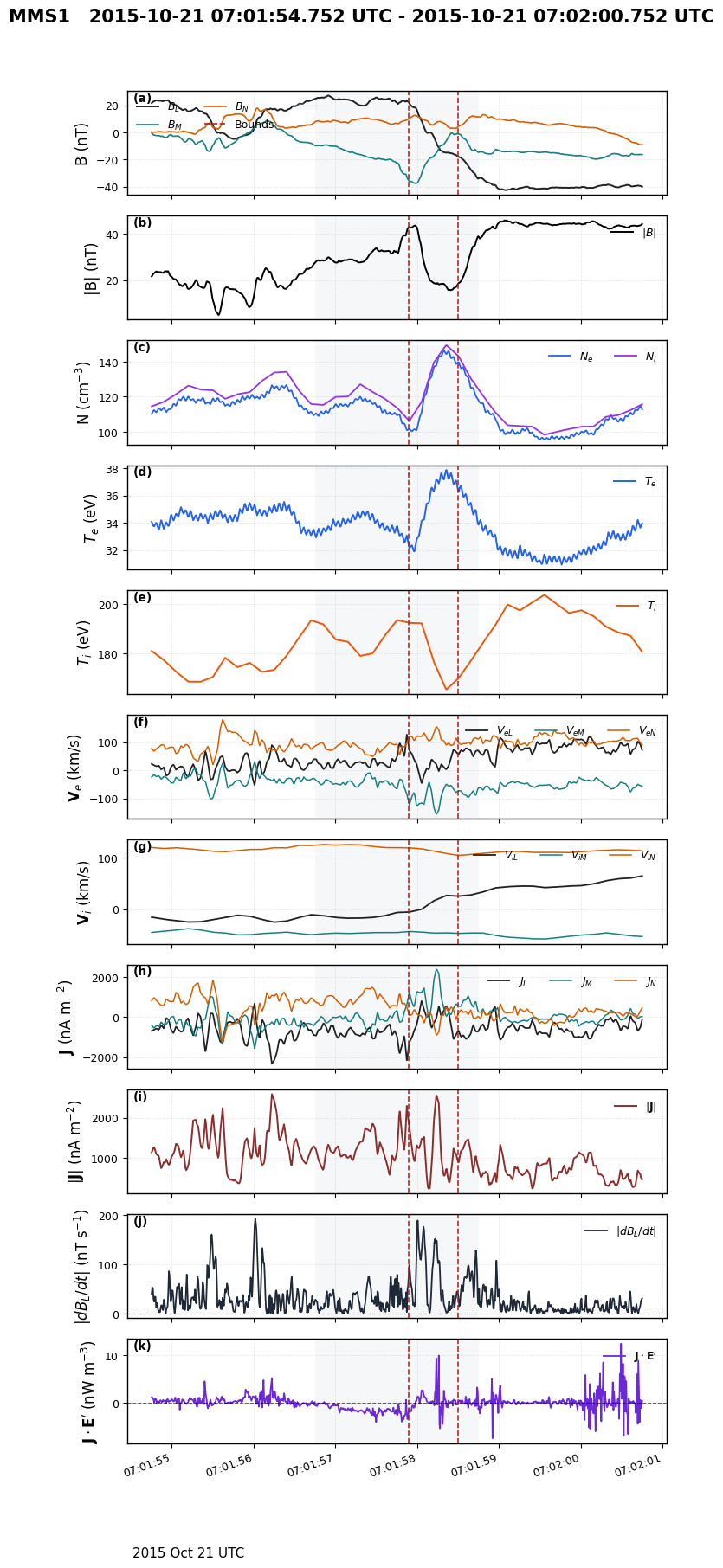}
\caption{Label-1 (candidate reconnection) detection in interval~1, 2015-10-21 07:01:57.900--58.500~UTC. Standard 11-panel detection plot with user-placed manual boundaries (red dashed) and the automated 2~s detection window (faded). Manual single-spacecraft workup: $\mathcal{M}=0.84$, $\Delta t_{cs}=0.60$~s, $|\vect{J}|_\text{peak}=2.56\;\mu\text{A/m}^{2}$, $\theta_{\text{shear}}=110^{\circ}$, thickness $145.7\,d_{e}$, $|\Delta u_{eL}|/|\Delta V_{A,L}|=0.92$. Passes the simple single-spacecraft jet-ratio check on the manual boundaries (Section~\ref{sec:method:manual}).}
\label{fig:good_p1}
\end{figure}

\begin{figure}[p]
\centering
\includegraphics[height=0.82\textheight,width=\textwidth,keepaspectratio]{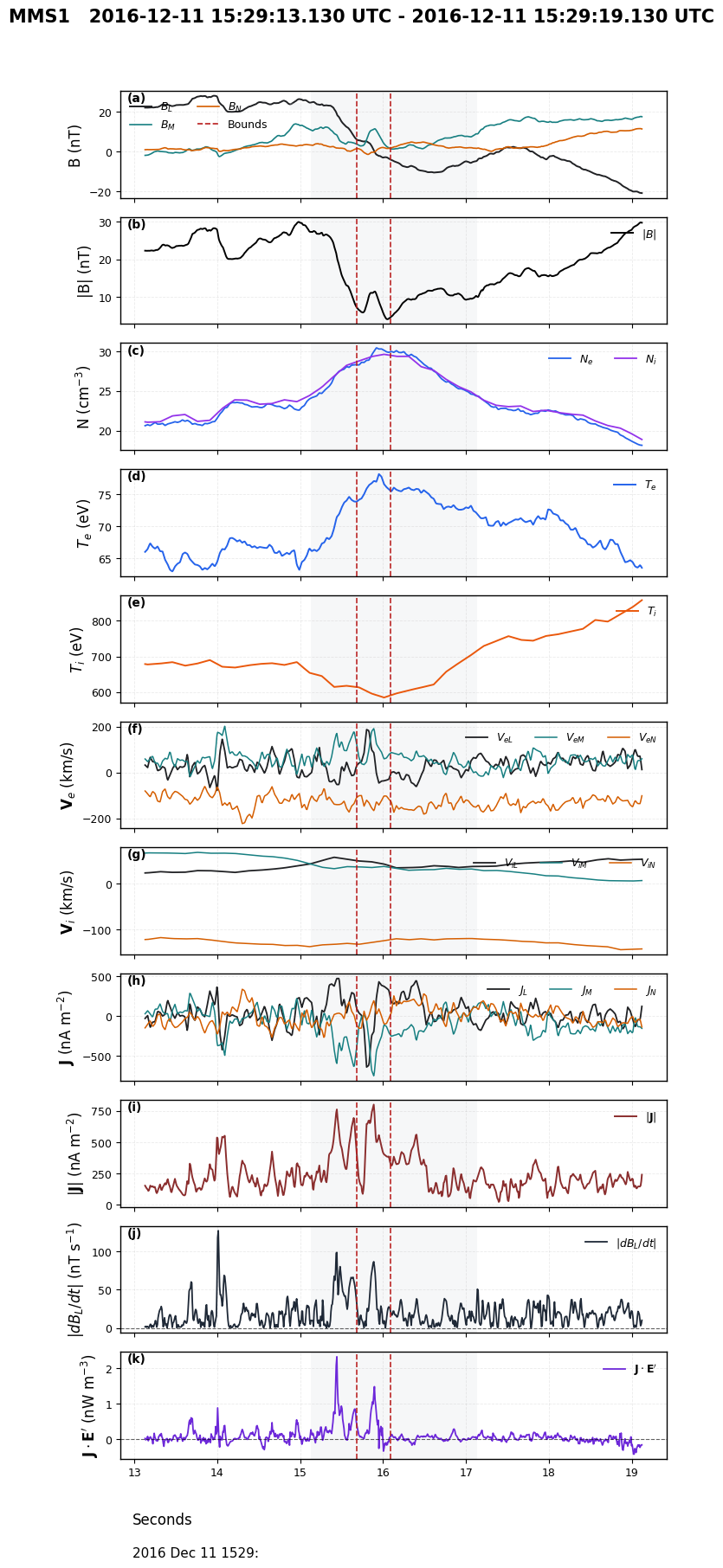}
\caption{Label-1 detection in interval~14, 2016-12-11 15:29:15.680--16.090~UTC. Plot conventions as in Figure~\ref{fig:good_p1}. Manual single-spacecraft workup: $\mathcal{M}=0.90$, $\Delta t_{cs}=0.41$~s, $|\vect{J}|_\text{peak}=0.80\;\mu\text{A/m}^{2}$, $\theta_{\text{shear}}=130^{\circ}$, thickness $53.5\,d_{e}$, $|\Delta u_{eL}|/|\Delta V_{A,L}|=1.29$.}
\label{fig:good_p2}
\end{figure}

\begin{figure}[p]
\centering
\includegraphics[height=0.82\textheight,width=\textwidth,keepaspectratio]{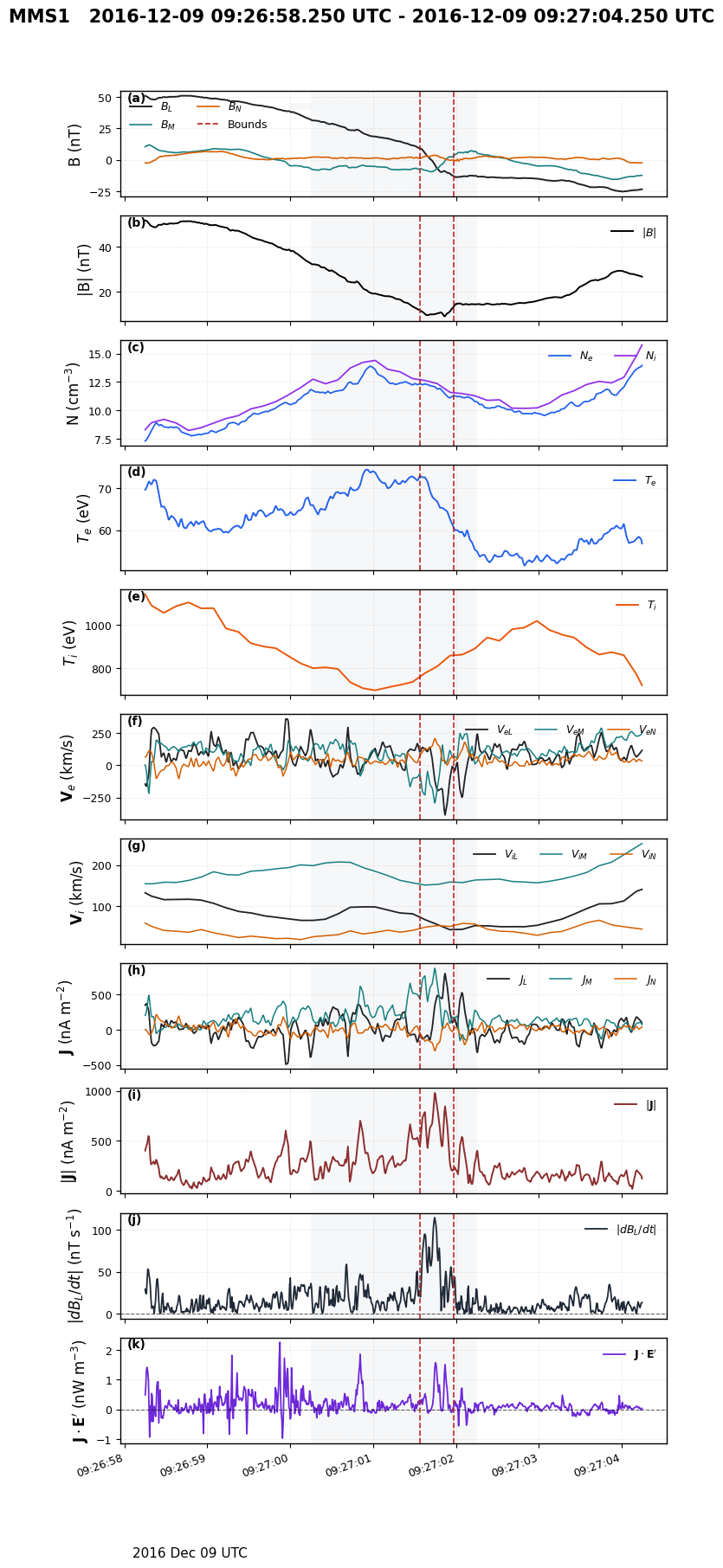}
\caption{Label-1 detection in interval~13, 2016-12-09 09:27:01.560--01.975~UTC. Plot conventions as in Figure~\ref{fig:good_p1}. Manual single-spacecraft workup: $\mathcal{M}=0.92$, $\Delta t_{cs}=0.42$~s, $|\vect{J}|_\text{peak}=0.98\;\mu\text{A/m}^{2}$, $\theta_{\text{shear}}=168^{\circ}$, thickness $10.6\,d_{e}$, $|\Delta u_{eL}|/|\Delta V_{A,L}|=0.97$.}
\label{fig:good_p3}
\end{figure}

Figures~\ref{fig:good_p1}, \ref{fig:good_p2}, and \ref{fig:good_p3} show three label-1 detections together with the numerical results of the manual single-spacecraft workup applied at user-placed boundaries. We discuss them in turn so that a reader can see what the workup numbers actually mean.

The first event (Figure~\ref{fig:good_p1}, interval~1, 2015-10-21 at 07:01:57.9 UTC) is the broadest of the three. The chosen reversal interval is $\Delta t_{cs} = 0.60$~s. The magnetic-field directions on the two sides of the interval differ by a shear angle of $\theta_{\text{shear}}=110^{\circ}$, and the peak in the single-spacecraft current proxy is $|\vect{J}|_\text{peak}=2.56\;\mu$A/m$^{2}$. Translating the duration into a spatial thickness through the iteratively converged ion-flow average gives $145.7\,d_{e}$, where $d_{e}$ is the electron skin depth, the characteristic scale of electron kinetic physics (electron demagnetization and the electron diffusion region); ion and electron dynamics themselves decouple at ion scales, the ion inertial length $d_{i}$ or the ion gyroradius depending on whether magnetic or electric-field dynamics are considered. Since $d_{i}\gg d_{e}$, thicknesses many tens of $d_{e}$ or more are effectively MHD-scale, while thicknesses near $\sim 10\,d_{e}$, well below $d_{i}$, correspond to the electron-only-reconnection regime. The present event is therefore mildly super-ion-scale, at the boundary of the MHD regime rather than deep within it. The single-spacecraft jet ratio is $|\Delta u_{eL}|/|\Delta V_{A,L}|=0.92$, meaning that the change in L-component electron-flow speed across the crossing matches the local change in Alfv\'en speed at the level of ten percent. Both parts of the simple decision rule of Section~\ref{sec:method:manual} are satisfied on this interval.

The second event (Figure~\ref{fig:good_p2}, interval~14, 2016-12-11 at 15:29:15.7 UTC) is narrower and crisper. The chosen reversal interval is $\Delta t_{cs}=0.41$~s, the shear is $\theta_{\text{shear}}=130^{\circ}$, and the current peak is smaller, $|\vect{J}|_\text{peak}=0.80\;\mu$A/m$^{2}$. The thickness in electron skin depths is $53.5\,d_{e}$, and the jet ratio is $|\Delta u_{eL}|/|\Delta V_{A,L}|=1.29$. A jet ratio above unity is consistent with an electron jet that is fully developed at the chosen boundary, which is the regime in which the simple jet test is most informative.

The third event (Figure~\ref{fig:good_p3}, interval~13, 2016-12-09 at 09:27:01.6 UTC) is the thinnest in the small sample, with $\Delta t_{cs}=0.42$~s, $\theta_{\text{shear}}=168^{\circ}$ (almost an antiparallel reversal), $|\vect{J}|_\text{peak}=0.98\;\mu$A/m$^{2}$, and a thickness of only $10.6\,d_{e}$. The jet ratio is $|\Delta u_{eL}|/|\Delta V_{A,L}|=0.97$. A thickness near $10\,d_{e}$ is in the range associated with electron-only reconnection in earlier MMS work \citep{Phan2018, Hubbert2022}, so this is the morphology in which the simple single-spacecraft test is closest to what one would expect from an active electron-only crossing.
\afterpage{\clearpage}

None of these three detections appears as an MMS1 reconnection event in the published catalog of \citet{Stawarz2022} over those intervals. We do not claim from a sample of three that any one of them is a confirmed reconnection event, because that confirmation would require multi-spacecraft validation and a more careful refinement of the crossing boundaries. We do claim that the pipeline surfaced sheet-like morphologies that survive a simple physical check on user-chosen boundaries and that a human reviewer found worth promoting to label~1 on the grounds defined in Section~\ref{sec:results:visual}. The working hypothesis is that a morphology-first one-class scorer is more permissive than peak-finding or threshold methods when the reconnecting sheet sits on top of an underlying turbulent structure, because the score depends on the multi-scale shape of the reversal and not on whether an isolated $|\vect{J}|$ peak rises far above the local background.

We tested this hypothesis empirically. For each of the 212 detections that are either sheet-or-better (label~1 or label~2) or overlap a published catalog event, we computed the ratio $\rho_{\mathbf{J}}\equiv\mathrm{median}(|\vect{J}|)/\max(|\vect{J}|)$ within the 2~s detection window; a low $\rho_{\mathbf{J}}$ corresponds to an isolated $|\vect{J}|$ peak above a quiet background, while a high $\rho_{\mathbf{J}}$ indicates a peak whose amplitude is comparable to surrounding fluctuations. The medians are $\rho_{\mathbf{J}}=0.32$ for the 22 catalog-matched detections, $0.37$ for the 75 non-overlap label-1 detections, and $0.38$ for the 115 non-overlap label-2 detections; the catalog-matched subgroup is significantly cleaner than each non-overlap subgroup by a one-sided Mann--Whitney $U$ test ($p\!=\!8.6\!\times\!10^{-3}$ against label~1, $p\!=\!2.8\!\times\!10^{-5}$ against label~2). Although the three medians are numerically close and the underlying distributions overlap substantially, the Mann--Whitney test compares the full distributions rather than the medians alone, and the significant $p$-values indicate a systematic, if modest, shift toward lower $\rho_{\mathbf{J}}$ in the catalog-matched subgroup rather than a large difference in central tendency. The same ordering holds when the analysis is repeated on $|dB_{L}/dt|$. This is consistent with the working hypothesis: the catalog-matched detections sit in the cleaner-peak regime where peak-amplitude criteria have clear contrast, and our additional candidates sit at systematically higher $\rho_{\mathbf{J}}$, where the peak/background contrast is reduced and a morphology-based scorer retains sensitivity that a peak-finder loses.

\begin{figure}[p]
\centering
\includegraphics[height=0.82\textheight,width=\textwidth,keepaspectratio]{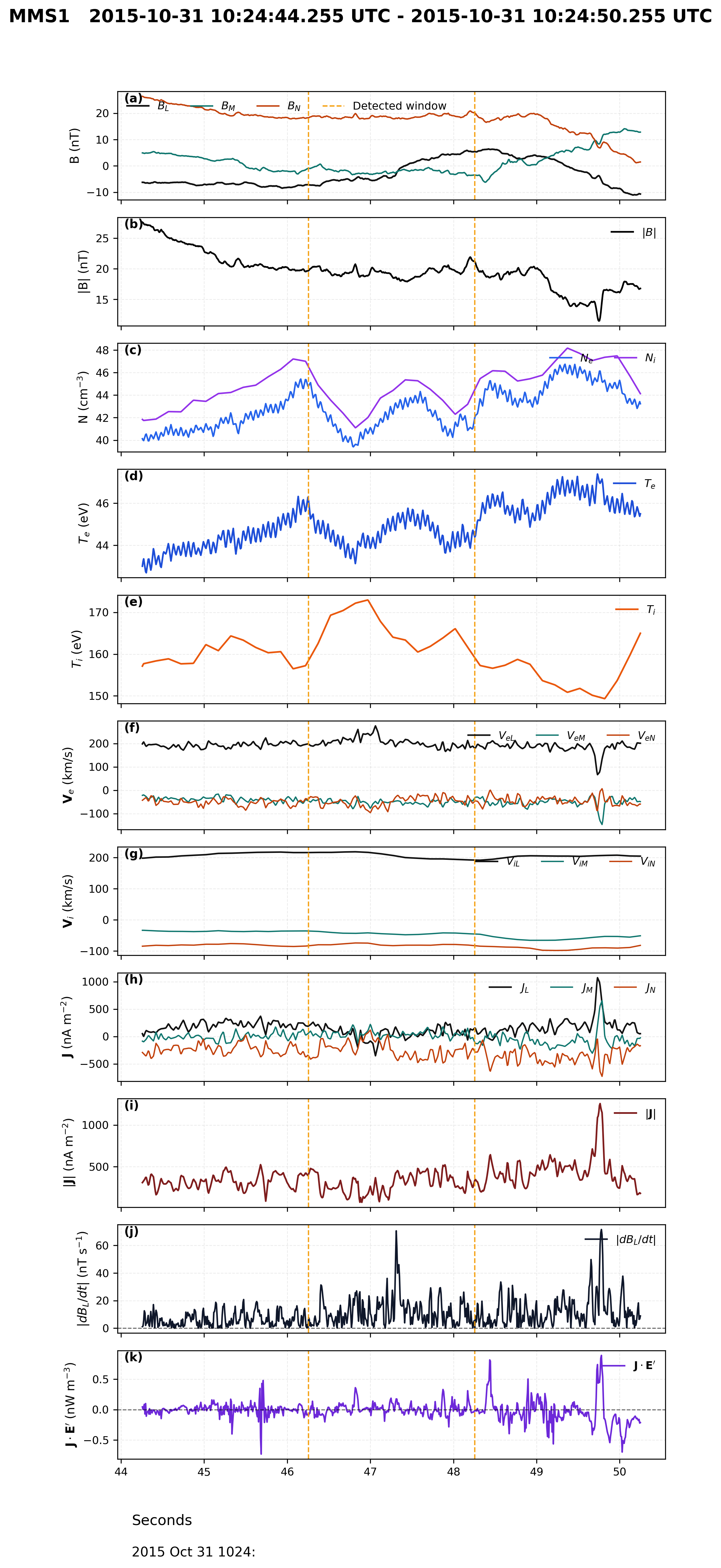}
\caption{Example of a label-3 (not-convincing) detection. The 2~s window is selected by the SVDD scorer because the magnetic channel contains a strong local gradient, but the co-located plasma response that distinguishes a current-sheet crossing is absent: $|\vect{J}|$ has no clean local enhancement aligned with the window, the electron and ion flow channels show no jet, and $\vect{J}\cdot\vect{E}^{\prime}$ is not robustly positive. Cases like this are caught at manual review, but they motivate the addition of flow-channel features to a future-version detector (Section~\ref{sec:discussion}).}
\label{fig:bad_example}
\end{figure}

Figure~\ref{fig:bad_example} shows the dominant failure mode of the pipeline. The selected 2~s window contains an SVDD-pass feature, but it is not a current-sheet crossing in the morphological sense the visual criterion requires. The magnetic component $B_{L}$ rolls but does not exhibit a clean sign reversal across a localised interval, and $|dB_{L}/dt|$ is elevated mainly because the trace is locally noisy rather than because of one sharp gradient at one specific time. The corroborating evidence that the visual criterion expects is missing on every channel that matters for reconnection. There is no clean local enhancement in $|\vect{J}|$ that lines up with the magnetic feature, the electron and ion flow channels do not show a localised L-component change that would correspond to a jet, and $\vect{J}\cdot\vect{E}^{\prime}$ is not robustly positive across the window. The reason this case slipped through the SVDD scorer is that the network sees only $B_{L}$ and $|dB_{L}/dt|$, and a window with strong derivative content can score high on those two channels alone without containing the plasma response that distinguishes a current-sheet crossing from a less interesting magnetic feature. The visual criterion catches these cases reliably (they fall into label~3), so the failure mode is contained at the manual review stage. It is, however, a clear motivation for adding flow-channel features to a future version of the detector, as discussed in Section~\ref{sec:discussion}.
\afterpage{\clearpage}

\section{Discussion}
\label{sec:discussion}

The headline outcome of the framework is operational rather than physical. A search space of \num{22775} sliding windows is too many to inspect by eye. A queue of 270 prioritised candidates is not. Converting a burst-mode dataset of order an hour into a tractable manual-review queue is what the pipeline is built to do, and the visual-screening result that 78\% of the queue ends up labeled sheet-like or reconnection-like indicates that the queue consists, for the most part, of windows that are worth looking at.

The same screening result tells us what the model is and is not doing. The mean SVDD score among label-1 detections (0.913) is indistinguishable from the mean among label-2 detections (0.922), a direct consequence of the training distribution: the model is trained against a library of sheet-like reversals whose multi-scale fingerprint matches a reconnecting reference event, so the score measures how sheet-like a window is, not how reconnecting it is. The same fact also explains why the label-2 population (sheet-like but not retained as reconnection-like) outnumbers the label-1 population by roughly 118 to 93. To separate the two populations the model would need additional features, plausibly the electron-jet channel, the $\vect{J}\cdot\vect{E}^{\prime}$ channel, or a local Wal\'en residual; adding any of those as in-class features, however, requires deciding what an out-of-class example looks like, which returns us to the exact problem that one-class detection was designed to avoid. A more promising route is a self-supervised second stage that operates on the present detector's output and uses the flow channels as additional features, listed in the future-work paragraph below.

The same division of labour explains how the present output relates to the published reconnection-event catalog over the same intervals. Of the 270 detections, 22 overlap a reconnection event in the catalog we cross-check against, a recall of 24/78 (30.8\%) of the catalog's own events over these intervals, and 18 of those 22 overlap detections were independently retained as label~1 in blind manual screening. The candidate-reconnection pool we surface is 93 detections in total, of which 18 are catalog-overlap and 75 are additional candidates that did not appear in the catalog over those intervals. This is consistent with the design of the two methods: the published catalog applies multi-spacecraft validation and a stricter event-finding logic and is therefore conservative by construction; the present pipeline is single-spacecraft and is deliberately tuned to surface a richer candidate pool for human review. We held the present configuration fixed throughout the 15-interval scan so that the result is reproducible end to end.

To our knowledge the framework is novel in three respects. First, anchoring the training distribution to a single well-characterised reference event, with a Monte Carlo generator producing the population from that one seed, sidesteps both the labeled-event-scarcity problem and the team-specific labeling-convention problem that supervised approaches inherit. Second, using $S_{2}(\tau)$ as the acceptance criterion ties the generator to a regime-specific multi-scale fingerprint rather than to an arbitrary similarity score. Third, framing the detection problem as morphology recognition rather than reconnection classification yields a Phase-1 candidate finder decoupled from the unresolved labeling debate, with the reconnection decision deferred to downstream multi-spacecraft review. The pipeline can be retargeted to a different magnetosheath regime by swapping the seed event and re-running the Monte Carlo and training steps; retargeting across reconnection classes additionally assumes that those classes differ in their magnetic-structure fingerprint as captured by $S_{2}(\tau)$, an assumption that is not tested here and would be more directly supported by extending the generator to additional feature channels, as discussed below.

The pipeline has three groups of limitations. \textit{First, downstream diagnostics inherit the chosen crossing boundary.} The jet ratio, the inflow Alfv\'en speed, and the thickness in $d_{e}$ can shift materially when the boundary is widened or narrowed by a fraction of a second \citep{Hasegawa2024}. Until boundary refinement is automated, the single-spacecraft jet-ratio check is a sanity test on a chosen interval rather than a substitute for four-spacecraft validation; the three label-1 detections in Figures~\ref{fig:good_p1}--\ref{fig:good_p3} pass that check but we do not promote them from candidate to confirmed on that basis. \textit{Second, the fixed configuration imposes scale constraints.} The 2~s scan window has reduced sensitivity to significantly narrower or broader structures, the 0.3~s merge gap implies a $\pm 1$~s tolerance when matching the retained midpoint window against a manually preferred crossing time, and the model sees only $B_{L}$ and $|dB_{L}/dt|$, enough to score sheet-like morphology but not to discriminate reconnecting from non-reconnecting sheets. \textit{Third, the single seed sets the regime of validity.} Regimes with systematically different fluctuation spectra (magnetopause, magnetotail, shock transitions) should not be expected to transfer without retraining on a regime-appropriate seed. The generator also assumes a quasi-one-dimensional crossing trajectory; different real crossings may sample the sheet at different angles, and turbulent variability is superimposed on the coherent structure. Because the acceptance test operates on shape rather than absolute duration, this primarily affects the inferred crossing width rather than the reversal profile itself, an effect already partly absorbed by drawing the half-thickness $w$ from a distribution rather than fixing it to a single value.

These limitations point to clear future work. The highest-priority follow-up is a multi-scale detector that scans more than one window length. A multi-seed library would broaden regime coverage, an automatic boundary-refinement stage would let Wal\'en-style analyses run without inheriting an arbitrary boundary, and a multi-spacecraft timing cross-check \citep{Dunlop2002, Vaivads2004} would close the gap between the present pipeline and curated multi-spacecraft catalogs. Adding electron-flow and energy-conversion channels as features, framed as a self-supervised refinement on the existing one-class output \citep{Ruff2020}, is the natural way to push the label-1-versus-label-2 discrimination into the automated stage. A complementary route that preserves the one-class framing rather than replacing it is to extend the Harris/$S_{2}(\tau)$ construction itself to these additional channels: symbolic regression \citep{Cranmer2020, UdrescuTegmark2020} could recover closed-form parametric templates for the electron- and ion-jet profiles or the $\vect{J}\cdot\vect{E}^{\prime}$ profile directly from the reference event, in the same spirit as the hand-chosen Harris ansatz for $B_{L}$, after which an analogous multi-scale acceptance band could be built around each recovered template to generate a richer surrogate library without requiring a curated negative class. This is a harder search problem than the present single-channel construction, since the additional channels are coupled and higher-dimensional rather than a single scalar reversal, but it offers a path to more discriminating features that does not reintroduce the labeled-negative-class problem the one-class design was built to avoid. Thin, short current-sheet crossings without a coupled ion jet, as characterised by \citet{Phan2018, Hubbert2022}, are the regime the present library is anchored around; the third example in Section~\ref{sec:results:examples} (Figure~\ref{fig:good_p3}, $10.6\,d_{e}$) sits closest to that regime.

\section{Conclusions}
\label{sec:conclusions}

We have presented a methods-oriented framework for narrowing the search for magnetic-reconnection candidates in MMS burst-mode data. The framework is built around three deliberate design choices: a single-event seed for the training distribution, an $S_{2}(\tau)$-anchored Monte Carlo surrogate library that supplies physically calibrated diversity around that seed, and a one-class Deep SVDD scorer placed downstream of a frame-quality LMN gate. On 15 magnetosheath turbulence intervals from the supplementary list of \citet{Stawarz2022} (1.58~h of burst-mode coverage) the pipeline compressed \num{22775} sliding windows to 270 final detections, a 98.8\% reduction. The principal output is the 93 candidate-reconnection detections that survive manual visual screening, with a further 118 sheet-like detections retained for follow-up; the candidate-reconnection pool includes all 18 detections that both overlap a published catalog event and pass the visual criterion. The contribution is to show that a Phase-1 candidate-finding stage can be built around one well-understood reference event and a physically calibrated surrogate library, that it compresses an MMS burst dataset to a manageable manual-review queue, and that the resulting candidate pool extends beyond what a single curated catalog records. Phase-2 problems (boundary refinement, multi-spacecraft validation, and a second-stage discrimination of reconnecting from non-reconnecting sheets) are the natural follow-up directions.

\section*{Acknowledgments}
We thank the MMS Science Team and the MMS Science Operations Center for the public release of MMS burst-mode data, and the pySPEDAS development team for the open-source data-access package used throughout this work \citep{Grimes2022}. We acknowledge the use of Magnetospheric Multiscale data accessed through the NASA MMS Science Data Center. This work was supported by the Partnership for Heliophysics and Space Environment Research (PHASER) cooperative agreement between NASA Goddard Space Flight Center and The Catholic University of America. Julia E. Stawarz is supported by the Royal Society University Research Fellowship URF\textbackslash R\textbackslash 251029. 

\section*{Open Research}
MMS burst-mode FGM, FPI, and EDP data used in this work are publicly available through the MMS Science Data Center (\url{https://lasp.colorado.edu/mms/sdc/public/}) and via the pySPEDAS package \citep{Grimes2022}. The 15 turbulence intervals and the published reconnection-event list are taken from the supplementary tables of \citet{Stawarz2022}.

\bibliographystyle{apacite}
\bibliography{references}

\end{document}